\documentstyle[preprint,pra,aps]{revtex}

\begin{document}

\bibliographystyle{prsty}

\tighten

\title{Quantum nondemolition measurement of a light field component by a
feedback compensated beam splitter}

\author{Holger F. Hofmann and Takayoshi Kobayashi}
\address{Department of Physics, Faculty of Science, University of Tokyo\\
7-3-1 Hongo, Bunkyo-ku, Tokyo113-0033, Japan}

\author{Akira Furusawa}
\address{Department of Applied Physics, Faculty of Engineering, 
University of Tokyo\\
7-3-1 Hongo, Bunkyo-ku, Tokyo113-8656, Japan}

\date{\today}

\maketitle

\begin{abstract}
In conventional quantum nondemolition measurements, the interaction
between signal and probe preserves the measured variable. Alternatively, 
it is possible to restore the original value of the variable by feedback.
In this paper, we describe a quantum nondemolition measurement of a 
quadrature component of the light field using a feedback compensated
beam splitter. The noise induced by the vacuum port of the beam splitter
is compensated by a linear feedback resulting in an effective
amplification of the observed variable. This amplification is then be 
reversed by optical parametric amplification to restore the original
value of the field component. 
\end{abstract}

\vspace{0.5cm}

The measurement backaction required by the uncertainty principle is a 
fundamental feature of quantum mechanics. In particular, this principle
can be applied in quantum communications to prevent or detect eavesdropping
\cite{Ral00}. 
However, most measurements in quantum optics are performed by 
irreversibly absorbing the measured light.
It is therefore a special challenge to devise experimental realizations 
of back-action evasion measurements that allow the non-destructive 
observation of light field properties \cite{Cav80}. 
Usually, quantum nondemolition measurements of the light field are realized by 
a non-linear coupling between the signal field and a meter field
\cite{Lev86,Fri92,Yur85,Por89,Per94}. In the following, a simple beam 
splitter will
be used to couple the signal and the meter fields. While this method initially
adds noise to the observed property of the signal field, it has been shown
that this noise can be compensated by a measurement dependent 
feedback \cite{Lam97}. The result of this compensation is an amplification of 
the observed field component. By reversing this amplification in an optical 
parametric amplifier, the original value of the observable is restored. 
It is then possible to realize a quantum nondemolition measurement of a 
quadrature component using only one single mode parametric amplification
step performed after the original beam splitter measurement. This approach thus
represents a considerable simplification compared to the previous realizations
of quadrature quantum nondemolition measurements which used two mode 
parametric amplification in order to couple the meter mode to the signal mode
\cite{Yur85,Por89,Per94}. 

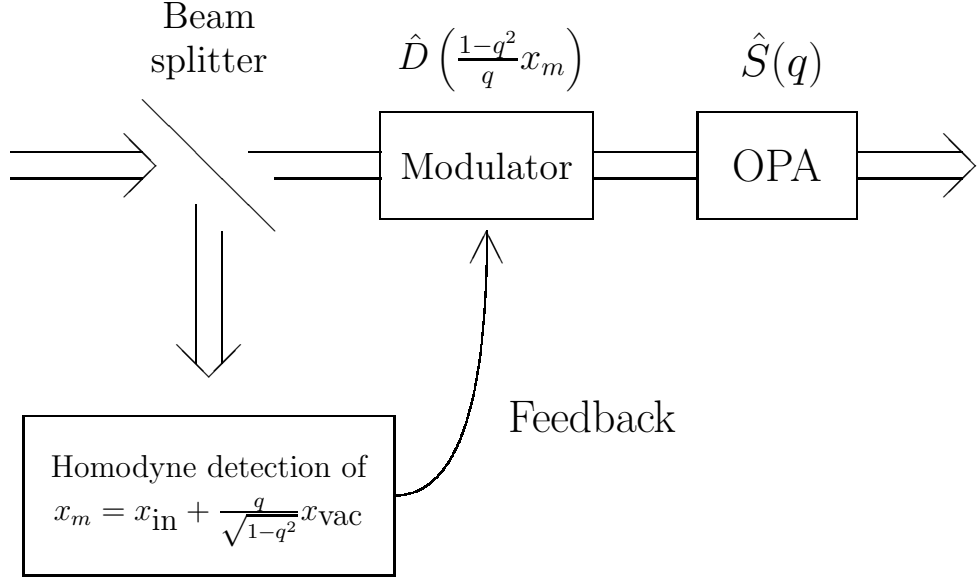
\begin{figure}

\begin{picture}(400,340)

\put(20,230){\line(1,0){50}}
\put(20,220){\line(1,0){50}}
\put(75,225){\line(-1,-1){12}}
\put(75,225){\line(-1,1){12}}

\put(70,250){\line(1,-1){50}}

\put(70,276){\makebox(50,12){\large Beam}} 
\put(70,258){\makebox(50,12){\large splitter}}

\put(110,230){\line(1,0){50}}
\put(120,220){\line(1,0){40}}
\put(160,205){\framebox(80,40){\large Modulator}}
\put(160,245){\makebox(80,40){
\large $\hat{D}\left(\frac{1-q^2}{q}x_m\right)$}}
\put(240,230){\line(1,0){40}}
\put(240,220){\line(1,0){40}}
\put(280,205){\framebox(60,40){\Large OPA}}
\put(280,245){\makebox(60,40){
\Large $\hat{S}(q)$}}
\put(340,230){\line(1,0){40}}
\put(340,220){\line(1,0){40}}
\put(385,225){\line(-1,-1){12}}
\put(385,225){\line(-1,1){12}}

\put(90,210){\line(0,-1){60}}
\put(100,200){\line(0,-1){50}}
\put(95,145){\line(-1,1){12}}
\put(95,145){\line(1,1){12}}

\put(25,70){\framebox(140,60){}}
\put(35,100){\makebox(120,20){Homodyne detection of}}
\put(35,80){\makebox(120,20)
{$x_m=x_{\mbox{in}}+\frac{q}{\sqrt{1-q^2}}x_{\mbox{vac}}$}}

\bezier{400}(165,100)(200,100)(200,200)
\put(200,200){\line(1,-2){6}}
\put(200,200){\line(-1,-2){6}}

\put(200,120){\makebox(80,20){\Large Feedback}}

\end{picture}

\caption{\label{setup} Schematic setup of the quantum nondemolition
measurement by feedback compensated beam splitter. The beam splitter 
reflectivity is $1-q^2$. The measured component $x_{\mbox{in}}$ mixes
with the vacuum component $x_{\mbox{vac}}$, but the feedback compensates
the contribution of $x_{\mbox{vac}}$ in the output field.}
\end{figure}

Figure \ref{setup} shows the experimental setup required for this quantum
nondemolition measurement. The quantum nondemolition measurement is 
composed of a sequence of three distinct steps, the beam splitter 
measurement, the linear feedback, and the parametric amplification.
In the first step, an unknown input state 
$\mid \psi_{\mbox{in}}\rangle$ is mixed with the vacuum at the beam splitter.
The product state of the input state and the vacuum can be expressed in terms
of the quadrature component $x_S$ of the signal mode and the quadrature 
component $x_M$ of the meter mode following the mixing of the modes at the
beam splitter described by the unitary transformation $\hat{U}_{BS}$. For
a reflectivity of $1-q^2$, the wavefunction of this quantum state reads
\begin{eqnarray}
\lefteqn{\langle x_S ; x_M \mid \hat{U}_{BS}\mid\psi_{\mbox{in}} 
; \mbox{vac.}\rangle =} 
\nonumber \\ &\hspace{1cm}& \left(\frac{\pi}{2}\right)^{1/4} 
\exp \left[ - (1-q^2)\left(x_S-\frac{q}{\sqrt{1-q^2}}x_M\right)^2\right]
\; \langle \sqrt{1-q^2}\, x_M + q x_S \mid \psi_{\mbox{in}}\rangle.
\end{eqnarray}
It is then possible to obtain a measurement result $x_m$ for the x-component
of the input field by measuring the meter variable $x_M$, rescaling the 
result such that $x_m = x_M/\sqrt{1-q^2}$. The non-normalized output state
in the signal field conditioned by this measurement reads
\begin{eqnarray}
\lefteqn{\mid \psi_{BS} (x_m) \rangle =}
\nonumber \\ &\hspace{0.5cm}& 
\left(\frac{2(1-q^2)}{\pi}\right)^{1/4} 
\! \int\! dx_S \; 
\exp\left[ - (1-q^2)\left(x_S - q x_m\right)^2\right]
\langle (1-q^2) x_m + q x_S \mid \psi_{\mbox{in}}\rangle \mid x_S \rangle.
\end{eqnarray}
The measurement probability is given by 
$P(x_m)=\langle \psi_{BS} (x_m) \mid \psi_{BS} (x_m) \rangle$.
The output amplitude of each component $\mid x_S \rangle$ now depends on the 
input amplitude at a x-value given by both $x_S$ and the measurement result. 

In the second step, the dependence of the output value of x on the noisy
measurement result is compensated by feedback. This feedback can be described 
by a displacement operator $\hat{D}(\Delta x)$ with
\begin{equation}
\Delta x = \frac{1-q^2}{q} x_m.
\end{equation}
The output state then reads
\begin{eqnarray}
\label{eq:amplify}
\hat{D}(\Delta x)\mid \psi_{BS} (x_m) \rangle &=& 
\left(\frac{2(1-q^2)}{\pi}\right)^{1/4} 
\! \int\! dx_S \;
\exp\left[ - \frac{1-q^2}{q^2}\left(q x_S - x_m\right)^2\right]
\langle q x_S \mid \psi_{\mbox{in}}\rangle \mid x_S \rangle.
\end{eqnarray}
The output component $\mid x_S \rangle$ is now associated with the
input amplitude of $\mid q x_S \rangle$. The vacuum noise component 
mixed into the signal by the beam splitter has been compensated 
completely, leaving only an effective amplification 
of the original field component $x_S$ by a factor of $1/q$.
This noiseless amplification has been realized experimentally 
by P.K. Lam and coworkers \cite{Lam97}. Equation (\ref{eq:amplify})
is a fully quantum mechanical formulation of this noiseless amplification
setup. 

In the final step of the quantum nondemolition measurement, 
the original value of $x_S$ is restored by 
a parametric amplification described by the squeezing operator 
$\hat{S}(q)$ with
\begin{equation}
\hat{S}(q)\mid x_S \rangle = \sqrt{q} \mid q x_S \rangle. 
\end{equation}
This operation attenuates $x_S$ back to its original value while 
simultaneously amplifying the conjugate quadrature component.
The output state of this sequence of beam splitter measurement, feedback 
displacement, and parametric amplification is then given by
\begin{eqnarray}
\label{eq:out}
\hat{S}(q) \hat{D}(\Delta x)\mid \psi_{BS} (x_m) \rangle &=&
\left(\frac{2(1-q^2)}{\pi \; q^2}\right)^{1/4} 
\! \int\! dx_S \;
\exp\left[ - \frac{1-q^2}{q^2}\left(x_S - x_m\right)^2\right]
\langle x_S \mid \psi_{\mbox{in}}\rangle \mid x_S \rangle.
\nonumber \\
\end{eqnarray}
Note that this state is not normalized, because the measurement probabilities
are given by $ P(x_m)=\langle \psi_{BS} (x_m) \mid \psi_{BS} (x_m) \rangle$.
Both the measurement statistics and the measurement back-action are thus 
described by equation (\ref{eq:out}). Since the output value of $x_S$
is now equal to the input value, the measurement may be described by an
operator commuting with the quadrature operator $\hat{x}_S$, such that 
\begin{equation}
\hat{S}(q) \hat{D}(\Delta x)\mid \psi_{BS} (x_m) \rangle = 
\hat{P}_{\delta\!x}(x_m) \mid \psi_{\mbox{in}}\rangle.
\end{equation}
The operator $\hat{P}_{\delta\!x}(x_m)$ is the generalized measurement 
operator for finite resolution quantum nondemolition measurements 
previously introduced in \cite{Hof00a,Hof00b}. Expressed in terms of the 
quadrature operator $\hat{x}_S$, it reads
\begin{eqnarray}
\label{eq:operator}
\hat{P}_{\delta\!x}(x_m) &=& \left(2 \pi \delta\!x^2\right)^{-1/4} 
\exp \left[-\frac{(x_m-\hat{x}_S)^2}{4\delta\!x^2}\right]
\nonumber \\
\mbox{with} && 4 \delta\!x^2 = \frac{q^2}{1-q^2}. 
\end{eqnarray}
The probability distribution over measurement results $P(x_m)$
and the normalized output state $\mid \psi_{\mbox{out}}(x_m)\rangle$
are then given by
\begin{eqnarray}
P(x_m) &=& \langle \psi_{\mbox{in}}\mid \hat{P}_{\delta\!x}^2(x_m) 
\mid \psi_{\mbox{in}}\rangle \\[0.2cm]
\mid \psi_{\mbox{out}}(x_m)\rangle &=& \frac{1}{\sqrt{P(x_m)}}
\hat{P}_{\delta\!x}(x_m) \mid \psi_{\mbox{in}}\rangle.
\end{eqnarray}
Note that the squared measurement resolution $4\delta\! x^2$ is given by the
ratio of the transmission and the reflectivity of the beam splitter.

Since the measurement back-action represented by $\hat{P}_{\delta\!x}(x_m)$
is the minimum required by the uncertainty principle,
the feedback compensated beam splitter represents an alternative
realization of optimal quantum optical tapping \cite{Ral00,Per94}.
The experimental effort should be greatly reduced by the use of a 
single mode parametric amplification instead of the two mode
amplification applied in \cite{Por89,Per94} according to the proposal of
\cite{Yur85}. Moreover, the formalism developed above allows an
assessment of a simple beam splitter measurement without feedback
and of noiseless amplification \cite{Lam97} in terms of an operator
product of the measurement operator $\hat{P}_{\delta\!x}(x_m)$ and the
corresponding unitary transformations,
\begin{eqnarray}
\label{eq:bsm}
\lefteqn{\hspace{-6cm}\mbox{Beam splitter measurement:}} \nonumber \\[0.2cm]
\mid \psi_{BS} (x_m) \rangle \hspace{0.5cm} &=& 
\hat{D}(-\Delta x)\hat{S}(1/q)
\hat{P}_{\delta\!x}(x_m) \mid \psi_{\mbox{in}}\rangle
\\[0.3cm]
\lefteqn{\hspace{-6cm}\mbox{Noiseless amplification:}} \nonumber \\[0.2cm]
\hat{D}(\Delta x) \mid \psi_{BS} (x_m) \rangle &=& \hspace{0.5cm}
\hat{S}(1/q)\hat{P}_{\delta\!x}(x_m) \mid \psi_{\mbox{in}}\rangle.
\end{eqnarray}
Every step of the quantum nondemolition measurement can thus be described 
by its own operator. This allows an in depth analysis of the information
and noise dynamics in quantum measurements. 
In particular, it should be
noted that equation (\ref{eq:bsm}) represents the losses of the beam 
splitter without feedback in terms of a measurement dependent unitary 
transformation. The intensity loss caused by the beam splitter seems to
originate from the dependence of the displacement operator 
on the classical information $x_m$.  This indicates that the loss in intensity
at the beam splitter is directly related to the classical information
gain represented by the measurement result $x_m$. 
The implications for photon losses can 
be investigated by applying the operator formalism to single photon states 
or to low amplitude coherent states. The efficiency of photon loss 
compensation may be relevant for implementations of eavesdropping strategies 
for single photon quantum communication. 
An investigation of photon loss compensation should also provide practical 
insights into non-classical photon-field correlations\cite{Hof00b}, since 
the measurement only registers the reflected field amplitude, not the 
photon losses. By performing photon counting
measurements on the transmitted signal field to monitor the changes
in photon number, a simplified version of the experiment proposed in 
\cite{Hof00b} can be realized.

In conclusion, we have described the realization of a quantum nondemolition
measurement of a quadrature component by a feedback compensated beam splitter
setup. This setup is more simple than previous realizations, since it requires
only a single mode parametric amplification. The operator formalism 
describing the three sequential steps allows a theoretical investigation
of the information dynamics in the quantum nondemolition measurement,
in a beam splitter without feedback, and in noiseless amplification.
Both the proposal for experiment and the formalism for its description 
should thus provide a helpful tool for the development of quantum 
communication technologies.



\end{document}